\documentclass{aa}
\usepackage{graphicx}
\newcommand{\teff}{T_{\rm eff}}
\newcommand{\mbol}{M_{\mbox{\rm bol}}}
\newcommand{\lsun}{L_{\mbox{\rm \sun}}}
\newcommand{\msun}{{M}_{\mbox{\rm \sun}}}
\newcommand{\rsun}{{R}_{\mbox{\rm \sun}}}

\newcommand{\aua}[2]{A\&A, #1 #2}
\newcommand{\auas}[2]{A\&AS, #1 #2}

\newcommand{\ha}{H$\alpha$}
\newcommand{\hb}{H$\beta$}
\newcommand{\hg}{H$\gamma$}
\newcommand{\hd}{H$\delta$}

\newcommand{\HeI}{He\,{\sc i}}

\newcommand{\FeI}{Fe\,{\sc i}}
\newcommand{\FeII}{Fe\,{\sc ii}}

\newcommand{\MgII}{Mg\,{\sc ii}}

\newcommand{\CaI}{Ca\,{\sc i}}

\newcommand{\OI}{O\,{\sc i}}

\newcommand{\SiIII}{Si\,{\sc iii}}

\newcommand{\CII}{C\,{\sc ii}}

\newcommand{\LiI}{Li\,{\sc i}}
\newcommand{\TiI}{Li\,{\sc i}}

\newcommand{\kms}{km\,s$^{-1}$}

\begin{document}
   \title{New spectroscopic observations of the B[e]/K binary system MWC\,623
   }

   \author{F.-J. Zickgraf
          }
   \institute{Hamburger Sternwarte, Gojenbergsweg 112, 21029 Hamburg, Germany}
   \offprints{F.-J. Zickgraf}
   \date{Received date; accepted date}

   \abstract{The B[e]/K binary system MWC\,623 was reinvestigated using new 
spectroscopic observations. The absorption lines of the K and the B star
do not exhibit any significant radial velocity variations over a time interval of 
14 years. The spectral classification using a recent echelle spectrum yielded spectral 
types of K2II-Ib and B4III. The luminosity class of the K star gives an estimate 
of the distance towards MWC\,623 of $2.4^{+1.4}_{-0.9}$\,kpc.
This is consistent with the kinematic distance of $2.0^{+0.6}_{-0.3}$\,kpc. 
The  masses derived from the locations of the binary components in the H-R diagram 
are $7\pm1.5$\,$\msun$ and $7.5\pm2.5$\,$\msun$ for the B and K star, respectively, i.e.
the mass ratio is close to 1. 
Both stars are coeval with an 
age of $50^{+10}_{-20}$\,Myr as shown by the comparison with isochrones. 
The high luminosity of the K star excludes a pre-main sequence evolutionary phase as 
explanation for the strong \LiI$\lambda$6708 absorption line observed in the late-type
component. Rather, the high 
lithium abundance is a consequence of the young age. Likewise, the B[e] star is a 
slightly evolved object starting it post-main sequence evolution.  
   }
\titlerunning{The B[e]/K binary system MWC\,623}
\authorrunning{F.-J. Zickgraf}
   \maketitle

\section{Introduction}
The emission-line object \object{MWC\,623} was described by Allen \& Swings 
(\cite{AS76}) as a peculiar Be star with a strong near-infrared excess. The 
spectrum they discussed was domi\-nated by strong emission lines of the 
Balmer series and numerous permitted and forbidden emission 
lines of \FeII , [\FeII ], and [\OI ]. Absorption lines were not reported. 
Zickgraf \& Stahl 
(\cite{ZickgrafStahl89}) (hereafter referred to as
Paper I) obtained high resolution spectroscopic observations 
of this object. In addition to the rich emission
line spectrum they detected early-type and late-type
absorption features, namely \HeI\ absorption lines and numerous
absorption features of \FeI , \TiI, \CaI , etc.. The analysis of  
these spectroscopic properties and of the continuum 
energy distribution lead Zickgraf \& Stahl to the conclusion that 
\object{MWC\,623} is actually a binary system consisting of a 
B2- and K2-type component. 

Polarimetry of \object{MWC\,623}  carried out by Zickgraf \& Schulte-Ladbeck 
(\cite{ZS89}) revealed 
the presence of intrinsic polarization. These observations were
interpreted as being due to a bipolar wind structure as suggested for B[e]-type
stars e.g. by Zickgraf et al. (\cite{Zickgrafetal85}) (cf. also the review by
Zickgraf, \cite{Zickgraf98}, on B[e]-type stars). The presence of intrinsic
polarization was considered indicative for an inclination angle deviating
significantly from pole-on.

A puzzling discovery was the existence of a strong absorption line of lithium,
\LiI$\lambda$6708\AA . Because it seemed unlikely that the late component in 
\object{MWC\,623} is a T\,Tauri star Zickgraf \& Stahl proposed  
that the high lithium abundance
of the K star  might be produced during He-flashes occuring when the 
K star evolves along the giant branch or alternatively that the object 
is just not old enough to have the lithium completely depleted. 

In this paper new spectroscopic observations of \object{MWC\,623} are presented. 
The purpose here is to investigate for the first time variability of the spectrum 
and to rediscuss the evolutionary scenario for MWC\,623 based on
a  revised spectral classification of the binary components. The
outline of the paper is as follows. 
The observations are described in Sect. \ref{obs}. 
Radial velocity measurements are discussed in Sect. \ref{rad}.
In Sect. \ref{var} spectroscopic
variability is investigated. The stellar para\-meters and evolutionary status 
are discussed in Sect. \ref{dis}. Finally, conclusions are summarized in Sect.
\ref{conclusion}

\section{New spectroscopic observations}
\label{obs}
MWC\,623 was observed on October 21, 1998 and June 16, 2000. 
The observations in
1998 were obtained with the spectrograph AURELIE at the 1.5\,m telescope of the
Observatoire de Haute Provence. A description of the
spectrograph can be found in Gillet et al. (\cite{Gilletetal94}). 
The spectra were observed with grating No. 2
with 1200\,lines\,mm$^{-1}$ giving a reciprocal linear dispersion of 8\,\AA\,mm$^{-1}$.
The detector was a double-barrette Thomson TH7832 (2048 pixel with 
13\,$\mu$m pixel size). The spectra cover the wavelength interval from
6540\,\AA\ to 6740\,\AA . The resolution of the spectra is 20\,000. Wavelength
calibration was obtained with Neon and Argon lamps.

The spectrum of June 2000 was observed with the echelle  spectrograph FOCES
(cf. Pfeiffer et al. \cite{Pfeifferetal98}) at the 2.2\,m telescope of  
Calar Alto Observatory. 
The spectrograph was coupled to the telescope with the
red fibre. The detector was a 1024$\times$1024 pixel Tektronix 
CCD chip with 24\,$\mu$m pixel size. With a diaphragm diameter of 
200\,$\mu$m and an entrance slit width of 180\,$\mu$m a spectral resolution of
34\,000 was achieved. Wavelength calibration was obtained with a ThAr lamp.

The spectra were reduced with the standard routines of the ESO-MIDAS software 
package (contexts {\em longslit} for the AURELIE spectra and {\em echelle}
for the FOCES data).
 
\section{Radial velocities}
\label{rad}

\begin{table*}
\caption[]{Heliocentric radial velocities of  emission and absorption lines 
between 1986 and 2000. The velocities of the emission lines for the years 
1986-88 are  from Paper I.}
\begin{tabular}{lllllllllll}
\noalign{\smallskip}
\hline
\noalign{\smallskip} 
date & \FeII\ em.& $[$\FeII ] & $[$\OI ]  & \ha &  \hb & \hg & \hd			&
\HeI\ abs.& \CII\ abs. & K star abs.   \\
1986 &$-5\pm2 $  & $-6\pm1 $ &$-5\pm4 $  &$+26\pm3$& &&$+8\pm3$ 			&
 $-4\pm4 $&$-8\pm3$&$+1\pm2 $\\
1987 & &$-8\pm2 $ &$-12\pm2 $&$+18\pm2$&& &						&	   & & $-4\pm2 $\\
1988 &$-6\pm4 $ & &$-6\pm4 $& $+22\pm2$&&&						&	    && $-3\pm2$ \\
1998 & & & &$+13\pm2 $&&&								&	    &&$+2\pm3 $\\
2000 & $-11\pm2 $ &  $-13\pm2 $& $-11\pm2 $&$+13\pm2 $ &$+11\pm2 $ &$+3\pm2$ & $-5\pm2$ &
 $-2\pm3 $& $-3\pm3$&$+0\pm2 $\\
\noalign{\smallskip}
\hline
\noalign{\smallskip}
\end{tabular}
\label{mwc623rv}
\end{table*}

\subsection{Absorption lines}
The radial velocity of the late-type component  
was determined from the spectra by means of a
cross-correlation technique, i.e. by measuring the shifts of the spectra 
relative to radial velocity
standard stars. For this purpose first the mean continuum was subtracted 
from the normalized
spectra. Then the spectra were rebinned on a
logarithmic wavelength scale. The relative shift was finally measured by
cross-correlation. The standard stars used had been observed 
with AURELIE during the run in October 1998. The selected stars were HD\,212943 (K0, $v_{\rm
rad} = 53.8$\,\kms ), HD186791 (K3, $v_{\rm rad} = -2.1$\,\kms ), and HD\,204867
(G0,  $v_{\rm rad} = +6.5$\,\kms ) (Duflot et al. \cite{Duflotetal95}).
In order to obtain consistent measurements of the radial velocities 
also for the spectra of 1986-88 discussed in Paper I, the same procedure was 
applied also to these data and the radial velocities were redetermined with the 
same set of standard star spectra.

The section of the spectra used for the measurement was 6600-6730\AA . This region
is completely dominated by the lines of the late-type component and is 
free of emission lines.
For the spectrum of 1987 a smaller interval of 6660-6710\AA\ had to
be used due the smaller wavelength interval covered by the spectrogram 
of this observing run (cf. Paper I). 
Measurement errors were estimated from the scatter of
the results for the three standard stars and from the scatter of the individual
spectra available for each observing campaign (2-3 spectra per observing run). 
Note, that only minor differences in the redetermined 
velocities  were found relative to the results given in Paper I.
 
The velocities of the \HeI\ absorption lines (and of the emission lines, see below) were
determined by  measuring directly the wavelengths instead of using the 
cross-correlation method. 

\begin{figure} 
\resizebox{\hsize}{!}{\includegraphics[angle=-90]{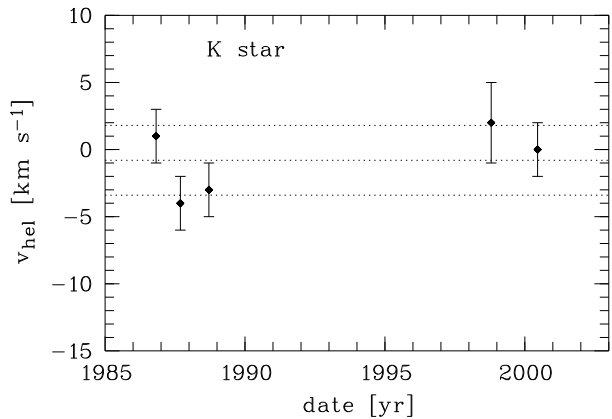}}
\caption[]{Heliocentric radial velocity of the K component. 
The interval of ${\sigma} = 2.6$\,\kms\ is marked by the upper and lower dashed lines. 
}
\label{mwc623kvel}
\end{figure}

In Table \ref{mwc623rv} the resulting  heliocentric radial velocities are
listed. The velocities measured from absorption lines of \FeI , \TiI , and 
\CaI\ are summarized as ``K star abs.''. 
Within the errors of the measurements no significant 
radial velocity variations of the absorption lines of the K star
could be detected between 1986 and 2000. This also holds for the \HeI\ lines 
attributed to the B star component.
In Fig. \ref{mwc623kvel} the heliocentric radial velocities of the
K star are plotted for the different observational epochs. 
The mean velocity of the K star calculated from the five measurements is
$v_{\rm K}= -0.8\pm2.6$\,\kms . The 
standard deviation of the mean velocity is of the
same order as the individual errors of the measurements. 

\subsection{Emission lines}

The radial velocities of the emission lines are also summarized in Table \ref{mwc623rv} including the
results from Paper I. 
In constrast to the absorption lines the emission features exhibit some variability.  
[\OI ] shows some variation on the order of 6\,\kms . In 2000 
the lines of \FeII\ and  [\FeII ]  had a slightly more negative velocity than in the 1980s. 
The mean difference is $\sim -6$\,\kms . Likewise, the Balmer lines in 1998 and 2000
exhibit a more negative velocity by the same amount as the singly ionized iron lines. 
The variable velocities of the Balmer lines 
could be related to line profile variations discussed in the next section.

\section{Spectroscopic variability}
\label{var}
\subsection{Emission lines}
\begin{figure}[tb]
\resizebox{\hsize}{!}{\includegraphics[angle=-90,bbllx=70pt,bblly=40pt,bburx=565pt,bbury=760pt,clip=]{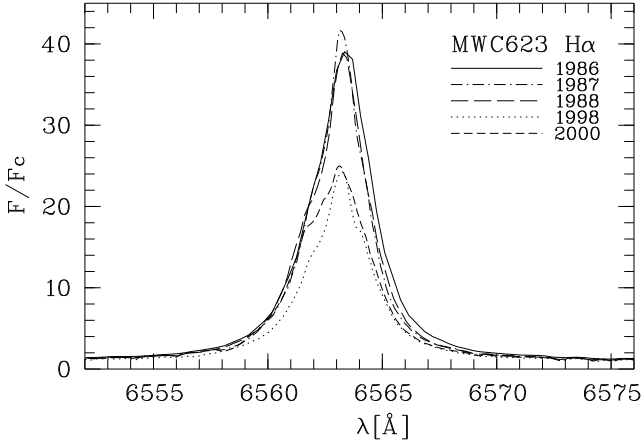}}
\caption[]{Line profile variations of \ha .  During 1986-1988 the emission line
was nearly twice a strong as in 1998 and 2000.
}
\label{mwc623haover}
\end{figure}

The emission line of \ha\ is clearly variable. This is shown in Fig. 
\ref{mwc623haover}. In the
recent spectra the line is a factor two weaker than in the late 1980s. Also small 
changes in the line profile are discernible, in particular in the blue wing which
showed a more pronounced dip in the 1998 and 2000 spectra. The spectrum of 1998
additionally shows a dip also on the red wing (weakest line in Fig. 
\ref{mwc623haover}). 

\begin{figure}[tb]
\resizebox{\hsize}{!}{\includegraphics[angle=-90,bbllx=70pt,bblly=40pt,bburx=565pt,bbury=760pt,clip=]{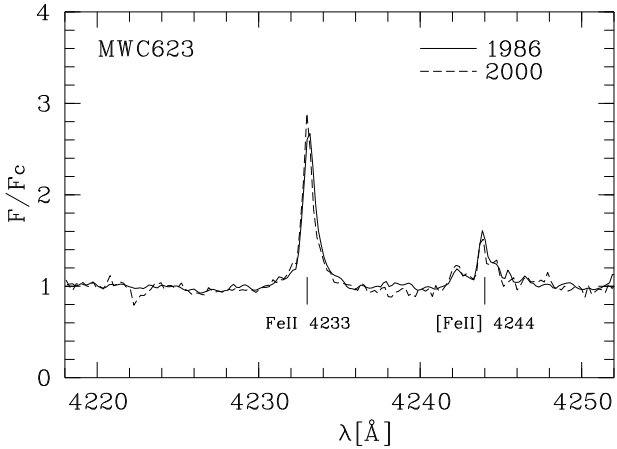}}
\caption[]{Comparison of the lines  \FeII\ and [\FeII ] in 1986 and 2000.  No significant variation
of the line strengths is discernible.
}
\label{mwc623feover}
\end{figure}
The emission lines of \FeII\  and [\FeII ], however, do not exhibit this 
variability pattern. 
Rather, these lines seem to be constant in intensity, at least for the two epochs of
1986 and 2000 (only these two observations of the blue spectral region are
available). This is depicted in Fig.
\ref{mwc623feover}. Note, that the spectrum of 2000 has been rebinned to the same
resolution as that of 1986 for this comparison.

\begin{table}
\caption[]{Equivalent widths $W$  of absorption lines around 6710\,\AA\ 
(given in m\AA ). The error of the individual measurements is of the order 
of 10\,m\AA .
The last column gives the sum of the equivalent width of the three \FeI\ lines
listed. Its error is about 20\,m\AA .}
\begin{tabular}{lllllll}
\noalign{\smallskip}
\hline
\noalign{\smallskip}
date   & \LiI     & \CaI &     \FeI    &     \FeI   &   \FeI   
& $\Sigma$ {\FeI} \\
       & $\lambda$6708         & $\lambda$6719     &  $\lambda$6703       &  $\lambda$6705 & $\lambda$6710 &  \\
\noalign{\smallskip}
\hline
\noalign{\smallskip}
1986 & 192      & 123     &  68               &    60             & 95  & 223 \\                      
1987 & 200      & 140     &   74              &    36             & 100 & 210\\
1988 &  298     & 142     &  67               &    62             &  88 & 217\\
1998 &  135     &  91     &   61              &    20             &  44 & 125   \\
2000 &  200     &  137    &   81              &    53             &  82  & 216   \\
\\
\noalign{\smallskip}
\hline
\noalign{\smallskip}
\end{tabular}
\label{mwc623li}
\end{table}

\subsection{Absorption lines}
\begin{figure}[tb]
\resizebox{\hsize}{!}{\includegraphics[bbllx=45pt,bblly=25pt,bburx=545pt,bbury=715pt,clip=]{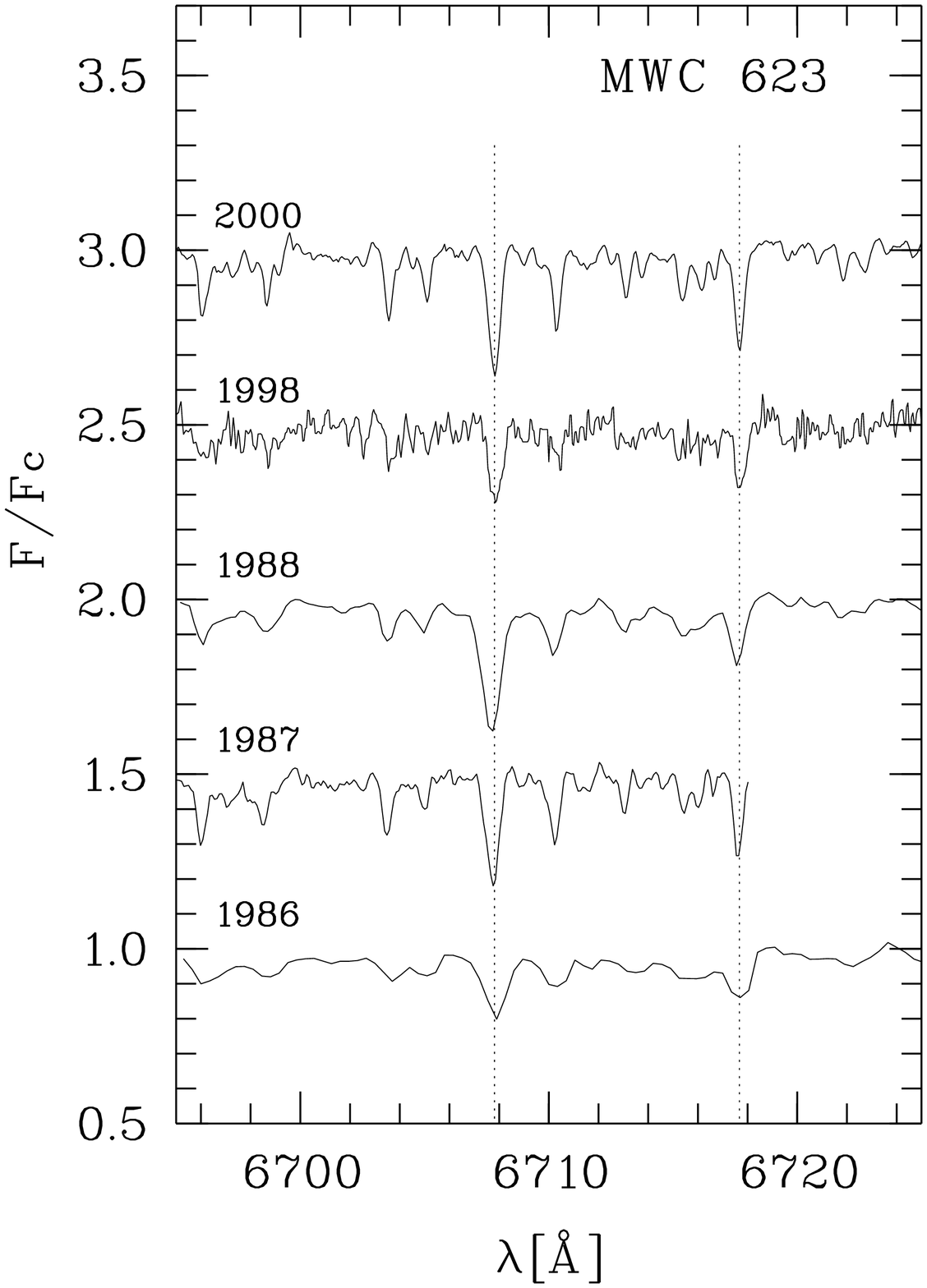}}
\caption[]{Sections of the spectrum around the absorption of 
\LiI$\lambda6708$\AA . The dashed line mark the positions of 
\LiI$\lambda6708$\AA\ and \CaI$\lambda6717$\AA . 
}
\label{mwc623liover}
\end{figure}

In  Fig. \ref{mwc623liover} a section of the spectrum of MWC\,623 around the line 
of \LiI$\lambda6708$\AA\ are shown for the five epochs of observation. 
Obviously, the strengths of the K star absorption features are variable to some
extent. The measured equivalent
widths $W$ of some characteristic lines of \FeI , of \LiI , and of \CaI\ are listed 
in Table \ref{mwc623li}.  In addition, the sum of the three \FeI\ lines is listed, which
has a smaller relative error than the individual lines.

\begin{figure} 
\resizebox{\hsize}{!}{\includegraphics[angle=-90,bbllx=40pt,bblly=40pt,bburx=565pt,bbury=745pt,clip=]{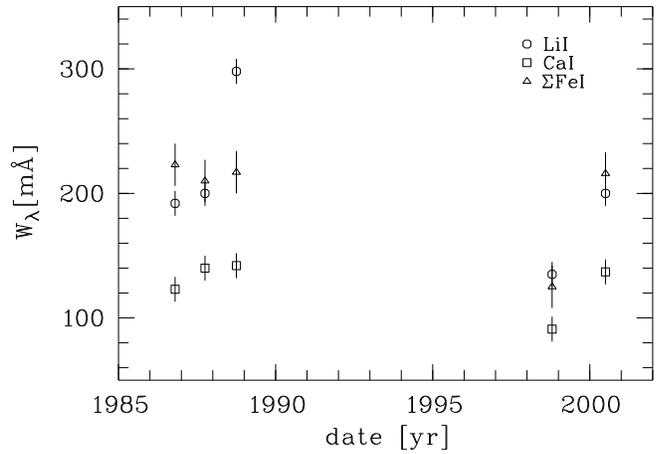}}
\caption[]{Variations of the equivalent widths of \LiI , \CaI , and the sum of the
three \FeI\ lines listed in Table \ref{mwc623li}. Note the weaker lines in 1998.}
\label{ploequi}
\end{figure}

Comparing the measured equivalent width in the different years which are 
plotted in Fig. \ref{ploequi} one realizes a
significant weakening of the strength of all lines in 1998 while in the other years
the line strengths are similar except the line of \LiI\ which  
shows a significant maximum in 1988. 

\begin{figure} 
\resizebox{\hsize}{!}{\includegraphics[angle=-90,bbllx=40pt,bblly=40pt,bburx=565pt,bbury=745pt,clip=]{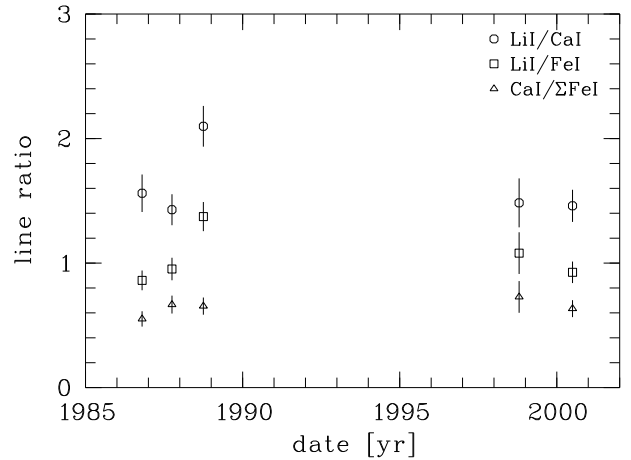}}
\caption[]{Ratio of the equivalent widths listed in Table \ref{mwc623li}.
Whereas the ratio of \CaI\ and \FeI\ is constant, even in 1998, a significant variation of 
the \LiI\ line is discernible in 1988 relative to  both,
\CaI\ and \FeI .}
\label{ploratio}
\end{figure}

The comparison of the relative line 
strength, i.e. $W$(\LiI )/$W$(\CaI), $W$(\LiI )/$W$($\Sigma$ \FeI), and 
$W($\CaI )/$W(\Sigma$ \FeI), however, does not reveal the drop in 1998. 
This is depicted in Fig. \ref{ploratio}. Rather, 
the line ratios are constant within the errors. However, 
the \LiI\ line again deviates significantly in 1988 in the sense of being stronger
than in the other years.

An explanation for the  variation of the measured equivalent widths 
in 1998 could be a varying relative brightness of the B and the K star. This 
would change the relative continuum level and consequently the equivalent widths of 
intrinsically constant lines. The weakening of the K star absorption lines in 
1998 would thus correspond either to a fading of the K star or to a brightening 
of the B star. Bergner et al. (\cite{Bergneretal95}) published $UBVRIJHK$ photometry for
MWC\,623 obtained in 1989-1994. 
Their mean $V$ magnitude was $10.89\pm0.06$ as compared to 
10.5 measured by Allen \& Swings (\cite{AS76}). This shows that MWC\,623 which is also
known as \object{V2028\,Cyg} is  photometrically variable at least on a 0.3-0.4 magnitude level 
in $V$. The mean $B-V$ colour of
Bergner et al. is $+1.21\pm0.05$ which is similar to +1.3 given by Swings 
(\cite{Swings81}). Unfortunately, the photometry available so far does not allow to
decide which of the components is variable although there might be some weak indication 
that the object appears bluer when it is weaker. This would suggest that 
the K star became fainter. 

The behaviour of the \LiI\ line remains puzzling. Whereas the
ratio of \CaI\ and \FeI\ is constant suggesting that no change of the effective
temperature of the K star has occured, the observation of 1988 seems to indicate a
real increase of the \LiI\ equivalent width although no explanation for such a 
behaviour can be give at this time. 
Certainly, further monitoring of the spectrum  is necessary in order to investigate 
this phenomenon in more detail.

\section{Discussion}
\label{dis}

\subsection{Distance towards MWC\,623 and stellar parameters}
\label{dist}
As discussed in Paper I the continuum energy 
distribution can be well fitted with the superposition of the energy flux 
distribution of a  B star, a K star, and a black body with $T_{\rm bb} 
\approx 930$\,K, assuming a reddening of $E_{B-V} = 0.8\pm0.2$. The spectral types 
adopted for the fit were B2V and K2III. The relative brightness of the components
lead to dereddened $V$ magnitudes of $V_0 = 8.6$ and 9 for the K and the B star,
respectively. The uncertainty of $V_0$  due to $E_{B-V}$ is  of the order of 0.6 mag.

In the following a new estimate of the spectral types of the components is obtained 
by making use of the wider spectral coverage provided by the FOCES echelle spectrum 
compared to the short spectral sections observed with the coud\'e spectrograph in 
1986-1988.

\begin{figure} 
\resizebox{\hsize}{!}{\includegraphics[angle=-90,bbllx=50pt,bblly=40pt,bburx=565pt,bbury=745pt,clip=]{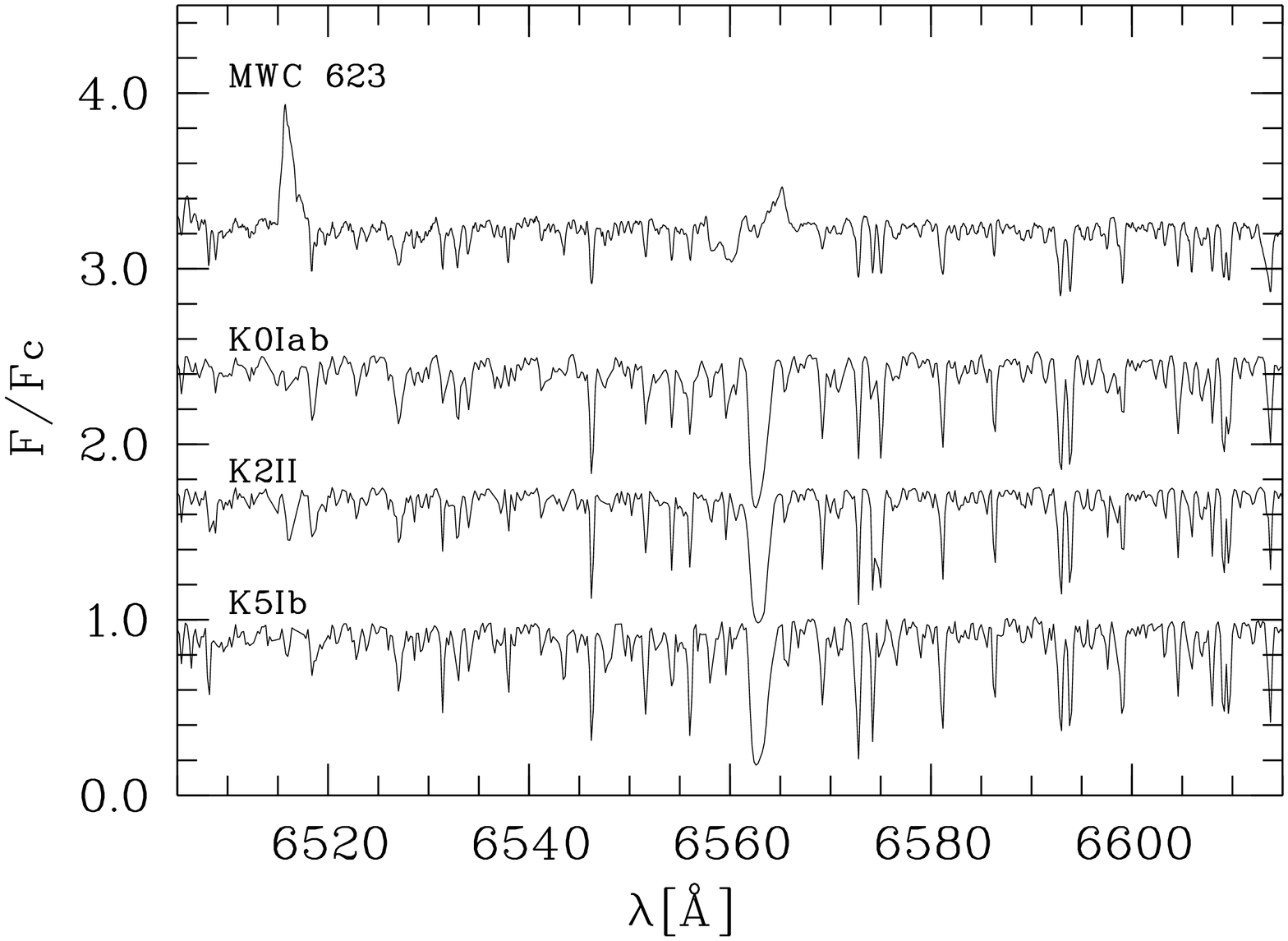}}
\resizebox{\hsize}{!}{\includegraphics[angle=-90,bbllx=50pt,bblly=40pt,bburx=565pt,bbury=745pt,clip=]{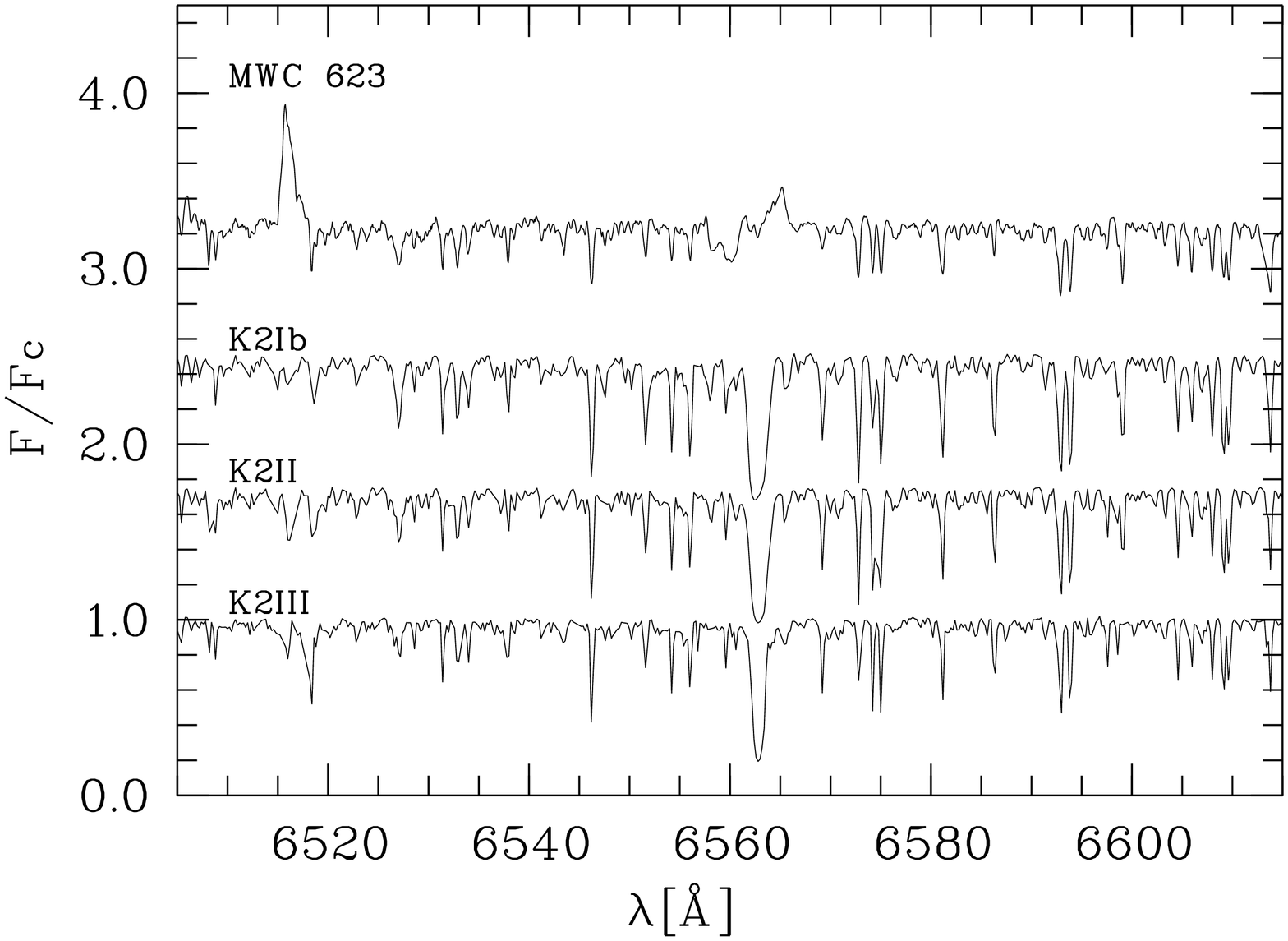}}
\caption[]{Spectrum of MWC\,623 together with spectra of stars with spectral types
between K0Iab and K5Ib (upper panel) and between K2III and K2Ib (lower panel). 
The \ha\ emission line of MWC\,623 has been removed by
normalization. The comparison shows good agreement of the K component of 
MWC\,623 with the K2II and K2Ib star. A spectral type of K2II-Ib is therefore 
adopted for MWC\,623. The K star spectra were obtained from the HYPERCAT data base of 
Prugniel \& Soubiran (\cite{PrugnielSoubiran01}).}
\label{ksternover}
\end{figure}

\subsubsection{Spectral types of the binary components}
The late-type component of MWC\,623 was re-classified by comparison with 
high resolution 
spectra of late G to K stars retrieved from  the stellar library 
of Prugniel \& Soubiran (\cite{PrugnielSoubiran01}). This library is part of the 
HYPERCAT\footnote{URL:
http://www-www-obs.univlyon1.fr/hypercat/11/\\spectrophotometry.html} data base. 
Note that the  resolution of the spectra in HYPERCAT is similar to that 
of the FOCES spectra. 

The comparison with stars of spectral types between G8 and K5
and of luminosity classes V to I yields the best agreement for spectral type K2II-Ib,
which will be adopted in the following. The uncertainty  of the spectral type
is about 1-2 subclasses.
Luminosity class V can be definitely excluded. Likewise, luminosity class III yields a
significantly poorer agreement than the higher luminosity spectra.
Fig. \ref{ksternover}  depicts the comparison of the late 
component of MWC\,623 with various K stars and with 
K2 stars of luminosity class III, II, and Ib 
(HD\,5234, HD\,39400, and HD\,206778, respectively).

\begin{figure} 
\resizebox{\hsize}{!}{\includegraphics[angle=-90,bbllx=80pt,bblly=65pt,bburx=555pt,bbury=770pt,clip=]{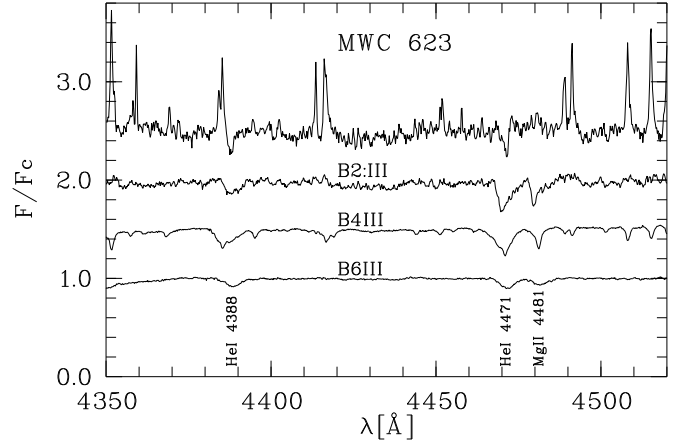}}
\caption[]{Spectrum of MWC\,623 together with spectra of stars with spectral types
between B2 and B6. 
Note, that in MWC\,623 \MgII$\lambda$4481  is in emission. 
The B star spectra were obtained from the stellar library of 
Prugniel \& Soubiran (\cite{PrugnielSoubiran01}).}
\label{heliumover}
\end{figure}

Likewise, the reclassification of the early-type component 
of MWC\,623  was aided by the comparison with B type 
spectra from the stellar library of Prugniel \& Soubiran. 
The classification of the late component suggests a luminosity class III for the 
hot component (see below).
A section around \HeI$\lambda$4388 and  \HeI$\lambda$4471 of the spectra 
is shown in Fig. \ref{heliumover}. The spectral type of the B component can be confined 
between B2 and B6 from the relative strengths of the \HeI\ lines,
by taking into account that the lines of \SiIII$\lambda\lambda$4553, 4568, 4575 
are absent,
and from the ratio of the very weak absorption feature at $\lambda4075$
and of  CII$\lambda4267$. 
In the following a spectral type of B4 will be adopted 
with an uncertainty of 2 subclasses. This is somewhat later than adopted in Paper I,
however, with only weak effects on the continuum fit as carried out in Paper I. 
In particular, the intrinsic colours are not much different and therefore the 
conclusions on the relative fluxes of the components and the dereddened magnitudes 
from Paper I are adopted also here in the following.

To summarize the result of the 
spectral classification it can be stated that the spectral
types of the  two components are  K2II-Ib and B4. 

\subsubsection{Kinematic distance estimate}
\begin{figure} 
\resizebox{\hsize}{!}{\includegraphics[angle=-90,bbllx=75pt,bblly=25pt,bburx=560pt,bbury=740pt,clip=]{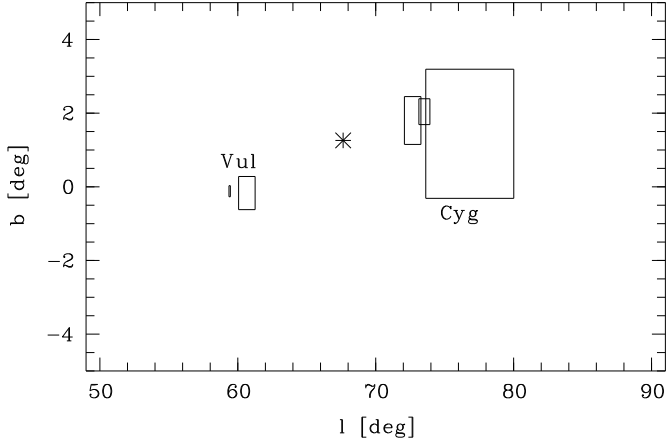}}
\caption[]{Position of MWC\,623 in the sky ($\star$ symbol) in galactic
coordinates and the locations of the (projected)
nearby associations Vul OB4, OB1, Cyg OB3\,A, B, and Cyg OB1.  
}
\label{mwc623assoc}
\end{figure}

As discussed in Paper I the projected position of MWC\,623 in the plane of sky 
at $l = 67\fdg64$ and $b = 1\fdg26$ 
lies between the Vul OB1, OB4 ($l \approx 60\degr$) and the Cyg OB3\,A, B and OB1
associations ($l \approx 73\degr$ and 76\degr, respectively). The 
distances of these associations are $d =$ 1.21, 2.54, 1.37, 1.82, and 2.31\,kpc
respectively. The association data were taken from 
Melnik \& Efremov (\cite{MelnikEfremov95}) who in addition to the location also
give the approximate extension of the associations. In Fig. \ref{mwc623assoc} 
the position of MWC\,623 in galactic coordinates 
and the associations located in the immediate (projected) neighborhood are plotted.  
MWC\,623 obviously is too far
from each of these associations to be  considered as a member of one of them.
Rather, MWC\,623 seems to be located somewhere in the field. Melnik \& Efremov also
give radial velocities for some associations, e.g. for Cyg OB3A, B, and OB1 
$-9$, $-17$, and $-6.5$\,\kms , respectively, however, with considerable scatter of 
7-10\,\kms\ rendering a comparison with MWC\,623 meaningless. 

An independent estimate for the distance may be obtained from the galactic 
rotation curve if it is assumed that the star is comoving with the field at its
given location in the Milky Way. 

\begin{table*}
\caption[]{Stellar parameters  of MWC\,623 for the 
distances $d = 2.4$\,kpc, corresponding to $m-M = 11.9$.
}
\begin{tabular}{llcclllll}
\noalign{\smallskip}
\hline
\noalign{\smallskip}
comp. & Sp. type & $\teff$ [K] &  $M_V$ & $\mbol$ & $\log L/\lsun$ & 
$R/\rsun$ & $M/\msun$ \\
\noalign{\smallskip}
\hline
\noalign{\smallskip}
B    & B4III    & 17200$\pm$3000 & $-2.9\pm1.0$ &  $-4.4\pm1.0$ & $3.7\pm0.4$ &  
 $8^{+11}_{-4}$ & $7.0\pm1.5$ \\
K    & K2Ib-II & 4300$\pm$200   & $-3.3\pm1.0$  & $-3.9\pm1.0$ & $3.5\pm0.4$ &  
$100^{+74}_{-42}$ & $7.5\pm2.5$ \\
\noalign{\smallskip}
\hline
\noalign{\smallskip}
\end{tabular}
\label{mwc623para}
\end{table*}

With a second order expansion of the galactic rotation curve the velocity 
of MWC\,623 in the local standard of rest  (LSR)
frame, $v_{\rm LSR}$, is given by
\begin{equation}
v_{\rm LSR} =  -2\,A\,\Delta R \sin l \cos b - 
  2\,A_2 (\Delta R)^2  \sin l \cos b
\label{rotation}
\end{equation}
where
\begin{equation}
v_{\rm LSR} = v_{\rm helio}+v_{\sun}
\end{equation}
 and
$\Delta R = R-R_0$, with $R$ and $R_0=8.5$\,kpc being the
galacto-centric distances of the star and the sun, repectively, $l$ and $b$ the
galactic longitudes and latitude, respectively, and the constants
$A=16.1$\,\kms\,kpc$^{-1}$ and
$A_2=-0.7$\,\kms\,kpc$^{-2}$ (cf. Dubath et al. \cite{Dubathetal88}). The solar
motion is given by
\begin{eqnarray}
v_{\sun}& = &
19.5\,{\rm km\,s}^{-1}\,(\cos \alpha_{\sun}\cos\delta_{\sun}\cos\alpha\cos\delta + \nonumber \\ 
& &\sin\alpha_{\sun}\cos\delta_{\sun}\sin\alpha\cos\delta+ 
\sin\delta_{\sun}\sin\delta)
\end{eqnarray}
where $\alpha$ and $\delta$ are the right ascension and declination of
the star, and $\alpha_{\sun}(1900)=18^{\rm h}$ and $\delta_{\sun}(1900)=+30\degr$ 
are the standard solar apex (Lang \cite{Lang80}).

The systemic velocity of MWC\,623 should, of course, be derived from 
the velocity of the centre-of-gravity of the binary system. However, the orbital
parameters of the system are still unknown. Therefore, as an 
estimate for the  centre-of-mass velocity the mean of the velocities of the 
photospheric lines of the B and the K
star in 1986 and 2000 was used, i.e. $v_{\rm helio} = -1.3\pm2$\,\kms . Note that the
mass estimates obtained below yielding a mass ratio of close to 1 
justifies this approximation.
In this way $v_{\rm LSR}
\approx +16.7$\,\kms \ with $v_{\sun} = +18$\,\kms , and
$\Delta R = -0.55$\,kpc is obtained from equ. \ref{rotation}. At the given galactic longitude two
solutions for the distance exist, $d_1 = 2.0^{+0.6}_{-0.3}$\,kpc and 
$d_2 = 4.4^{+0.4}_{-0.5}$\,kpc. 
The uncertainties given here result from the statistical error of the 
adopted radial velocity of 2\,\kms\ only. No systematic errors like peculiar
velocity were taken into account. The distances correspond to distance 
moduli of $m-M = 11.5^{+0.5}_{-0.4}$, and  $m-M = 13.2^{+0.2}_{-0.3}$, 
respectively. Note that both distances are consistent with the extinction 
$A_V \approx 1.9-3.1$ which in the direction towards MC\,623 occurs at distances
larger than 2-3\,kpc ( Neckel \& Klare \cite{NeckelKlare80}). 

\subsubsection{Distance and stellar parameters}
The spectral classification of the late component yielded a
spectral type of K2II-Ib which leads to an absolute visual magnitude of $M_V = -3.3\pm1$
(Schmidt-Kaler \cite{SK82}). With
the dereddened $V$ magnitude derived in Paper I of $V_0 = 8.6$ this yields $m-M =
11.9\pm1$ or $d = 2.4^{+1.4}_{-0.9}$\,kpc. Note the good agreement with the short 
kinematic distance $d_1$. 
The long kinematic distance of 4.4\,kpc  corresponds to $M_V = -4.6^{+0.2}_{-0.3}$ 
for the K star which would still
marginally be consistent with the spectroscopic distance, although the overall
agreement of spectroscopic and kinematic distance is better for the short distance.
Therefore in the following the spectroscopic distance of 2.4\,kpc is adopted.
The dereddened visual brightness of the early-type component of $V_0 = 9$ then 
leads to luminosity class III for the B star, i.e. a spectral type B4III. The 
stellar parameters of the components of MWC\,623 derived for $d = 2.4$\,kpc 
are summarized in Table \ref{mwc623para}. 
Bolometric corrections were taken from  Schmidt-Kaler (\cite{SK82}).

\begin{figure}[tb]
\resizebox{\hsize}{!}{\includegraphics[angle=-90,bbllx=190pt,bblly=50pt,bburx=550pt,bbury=545pt,clip=]{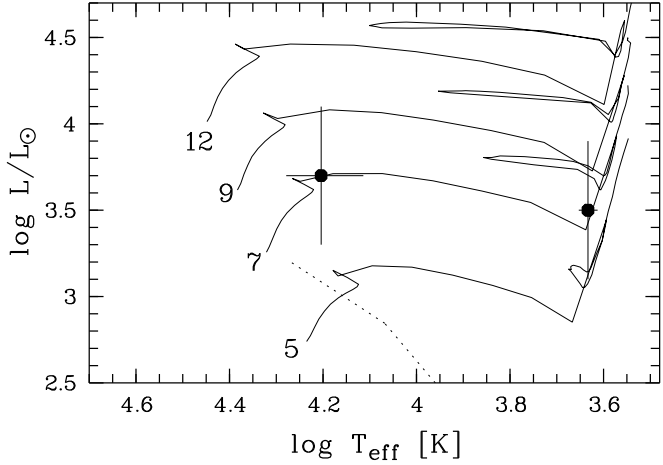}}
\caption[]{H-R diagram showing the locations of the B  
and the K component of MWC\,623. 
Evolutionary tracks are from Schaller et al. 
(\cite{Schalleretal92}) for metallicity $Z = 0.020$. ZAMS masses are indicated for
each track. The dashed line represents the stellar birth line from Palla \& Stahler
(\cite{PallaStahler93}). 
}
\label{mwc623hrd}
\end{figure}

In Fig. \ref{mwc623hrd} the location of the components of MWC\,623 on the H-R
diagram are shown. Evolutionary tracks were taken
from Schaller et al. (\cite{Schalleretal92}) for metallicity $Z = 0.020$. 
The masses estimated from the tracks are 6-9$\,\msun$ and 5-10$\,\msun$ for the B and the K star, respectively 
(see Table \ref{mwc623para}). The resulting mass ratio is $M_{\rm K}/M_{\rm B} =
1.07\pm0.4$.

\begin{figure}[tb]
\resizebox{\hsize}{!}{\includegraphics[angle=-90,bbllx=75pt,bblly=35pt,bburx=555pt,bbury=715pt,clip=]{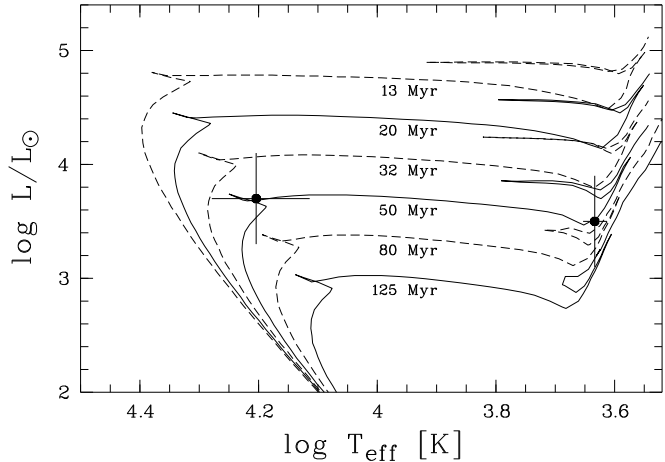}}
\caption[]{Isochrones  
from Bertelli et al. (\cite{Bertellietal94}) for metallicity 
$Z = 0.020$ with the positions of the B  
and the K component of MWC\,623. Ages are indicated for each track.
}
\label{mwc623iso}
\end{figure}

\subsection{Evolutionary status}
A possible model for MWC\,623 could involve interaction in a close binary system.
However, this
was already rejected in Paper I based on observations covering about 2 years. 
The absence of significant radial velocity variations of the photospheric absorption
lines of the B and the K star on an even longer timescale 
of 14 years as discussed in Sect. \ref{rad} indicates that the orbital period  
is actually on the order of longer than 14 years unless the small radial velocity 
variation is due to a low
inclination angle. This seems, however, to be inconsistent with
the observed polarization. Rather, the inclination angle
is expected to be intermediate to large.  

If the assumption is made that the scattering plane responsible for the observed
polarization is correlated with the orbital plane of the binary, the presence of strong 
intrinsic polarization suggests that the inclination angle $i$ deviates significantly 
from pole-on. Because the polarization is proportional to $\sin^2 i$ the 
inclination should be intermediate to large, i.e. $i$ should be 
of the order of $\ga$30-45\degr . Otherwise, the observed degree of 
intrinsic polarization, $p_{\rm obs} = p_{\rm true} \sin^2i$ of about
2\% (Zickgraf \& Schulte-Ladbeck \cite{ZS89}) would lead to unreasonably high true
polarization $p_{\rm true}$.

Assuming a minimum separation of the components on the order of the radius of the 
K star, i.e.100\,$\rsun$, the masses determined above would lead to a period of 
$\sim30$ days from 
 \begin{equation}
P = \frac{a^{3/2}}{M^{1/2}} \,{\rm yr}
\label{eqbinary}
\end{equation}
with the separation $a$ (in A.U.) and the total mass $M = M_1+M_2$ (in $\msun$)
of the system. In this case the velocity amplitude would be of the order of 80\,\kms . 
This is inconsistent with the observations because even within a few days of observation 
as in 1988 (cf. Paper I) significant velocity variations should have been detectable.
On the other hand, the small observed velocity variation $\Delta v\sin i \la 2-3$\,\kms\ over a time
interval of 14 years
suggests that the orbital period $P$ is longer than $\sim14$ years. 
This leads to  $a \ga 14$A.U. $\approx 3000\rsun$. 
The velocity amplitude would then be
of the order of $\la15$\kms . Taking $\sin i$ into account expected velocity variations
are thus on the order of at most a few \kms , which is consistent with the observations.

The linear separation of the components of 14 A.U. corresponds to an angular separation of 
$\sim5$\,mas. By taking into account the different colours of the components a positional shift should be  
directly measurable with the forthcoming astrometric space missions DIVA and GAIA  by 
comparing observations in different colours.

These estimates suggest that MWC\,623 is a wide binary system. 
In this case each component evolved like a single
star. From the evolutionary timescales 
for single stars it is expected that the B star is less  massive  
than the K star which then would evolve faster. This is in agreement with the 
mass estimates given above. 
Figure \ref{mwc623iso} shows the  H-R diagram with isochrones from Bertelli et
al. (\cite{Bertellietal94}) for metallicity 
$Z = 0.020$.  The positions of the B and K component 
confirm that both are coeval within the error bars. Both
stars lie  on the $50$\,Myr isochrone 
within an uncertainty range from $\sim30$ to $\sim125$\,Myr for the K star and 
$\sim30$ to $\sim60$\,Myr for the B star. The B component thus constraints the age
to $50^{+10}_{-20}$\,Myr.

These findings shed new light on the question of origin of the strong lithium line
in MWC\,623 which, in principle,  could indicate a pre-main sequence status for the K
star. However, the spectroscopic characteristics exclude a low-luminosity K star and
lead instead to a luminosity which is more than 2 orders of magnitudes above the stellar 
birth line (cf. Palla \& Stahler \cite{PallaStahler93}) which for $\teff = 4300$\,K 
is located at $\log L/\lsun \approx +1$. The K star is therefore clearly in a 
post-main sequence evolutionary phase.

The estimated age of 50\,Myr is younger than that of the Pleiades
(80-100\,Myr). The lithium abundance of the
late-type stars in this cluster is still high due to the young age.
The high lithium abundance of the late component of MWC\,623 can therefore quite
naturally be explained by the age in the sense that
the primordial lithium is not yet destroyed. 

The B[e] star in MWC\,623 is located slightly off but still close to the main 
sequence and  could thus be related to the class of (near) main-sequence B[e]-type 
stars discussed by Lamers et al. (\cite{Lamersetal98}) and Zickgraf 
(\cite{Zickgraf98}) for which  \object{HD\,45677} and \object{HD\,50138} are examples.
For this type of B[e] star it is unclear whether they are still in a pre-main
sequence phase of evolution  
or if they are already in the core-hydrogen burning phase. The first possiblity 
would place them on contraction tracks of Herbig
Ae/Be stars towards the main-sequence instead of post-ZAMS evolutionary tracks.
For  MWC\,623 is was possible for the first time to obtain a reliable age estimate.
It clearly excludes the pre-main sequence option at least for 
this object. Note also that both components of MWC\,623 are located above the
stellar birth line (cf. Fig. \ref{mwc623hrd}). This is also inconstent with a
pre-main sequence status.  The conclusion is that the B[e] component in MWC\,623 is 
a slightly evolved object starting its post-main sequence evolution. The similarities 
with  the mentioned (near) main-sequence B[e]-type stars 
suggest that this conclusion  might also hold for the latter class. 

The B[e] component in MWC\,623 might even belong to 
the subgroup of post-main sequence B[e] stars discovered in the Magellanic 
Clouds at luminosities $\log L/\lsun \approx 4$ (Gummersbach et al. 
\cite{Gummersbachetal95}). These objects are located in the H-R diagram 
below the luminous B[e] supergiants
described by Zickgraf et al. (\cite{Zickgrafetal86}) and extend the class of 
B[e] supergiants towards lower luminosities into the range of classical Be stars.

\section{Conclusions}
\label{conclusion}
New spectroscopic observations were used  to reinvestigate the B[e]/K binary system MWC\,623. Neither 
the K nor the B star absorption lines were found to exhibit significant radial velocity variations 
over a time interval of 14 years suggesting that the orbital period is longer than 14 years. 

The spectral classification using a recent echelle spectrum yielded spectral types of 
K2II-Ib and B4III. The luminosity class of the K star was used to estimate the distance 
towards MWC\,623 to 2.4\,kpc which is consistent with the kinematic distance of 2.0\,kpc. 

Placing the binary components in the H-R diagram allows to derive masses of 7 and 7.5\,$\msun$ for the B and K star,
respectively, yielding a mass ratio close to 1. 
Both stars are coeval with an age of 50\,Myr as shown by the comparison with isochrones. 

The high luminosity of the K star excludes a pre-main sequence evolutionary phase as explanation for the
strong \LiI$\lambda$6708 absorption line. Rather, the high lithium abundance is a consequence of the young
age.  Likewise, a pre-main sequence Herbig Ae/Be  phase of the B[e] star can be excluded.
Instead  it is  a slightly evolved post-main sequence object. 

In order to determine the orbital parameters of MWC\,623 and to investigate in more
detail the variablility indicated by the presently available observations it is
clearly desirable to monitor this object spectroscopically and photometrically also 
in the future.
 
\acknowledgements{This work made extensive use of the 
stellar library of high-resolution spectra by C. Soubiran and P. Prugniel 
available in the HYPERCAT data base via the World Wide Web. 
}


\begin{thebibliography}{}

\bibitem[1976]{AS76}
Allen D.A., Swings J.P., 1976, \aua{47, 293}

\bibitem[1995]{Bergneretal95}
Bergner Yu.K., Miroshnichenko A.S., Yudin R.V., Kuratov K.S., Mukanov D.B., Shejkina
T.A., 1995, A\&AS, 112, 221

\bibitem[1994]{Bertellietal94}
Bertelli G., Bressan A., Chiosi C., Fagotto F., Nasi E., 1994, A\&AS, 106, 275

\bibitem[1988]{Dubathetal88}
Dubath P., Mayor M., Burki G., 1988, \aua{205, 77}

\bibitem[1995]{Duflotetal95}
Duflot M., Figon P., Meyssonnier N., 1995, A\&AS, 114, 269

\bibitem[1994]{Gilletetal94}
Gillet D., Burnage R., Kohler, D., 1994, A\&AS, 108,181 

\bibitem[1995]{Gummersbachetal95}
Gummersbach C.A., Zickgraf F.-J., Wolf B., 1995, A\&A, 302, 409 

\bibitem[1998]{Lamersetal98}
Lamers H.J.G.L.M., Zickgraf F.-J.,  de Winter D., Houziaux L., J. Zorec J.,
1998, A\&A, 340, 117

\bibitem[1980]{Lang80}
Lang K.R, 1980. {\it Astrophysical Formulae}, Springer Verlag, Berlin,
Heidelberg, New York

\bibitem[1995]{MelnikEfremov95}
Melnik A.M., Efremov Yu.N., 1995,  Pis'ma Astron. Zh., 21, 13 

\bibitem[1980]{NeckelKlare80}
Neckel Th., Klare G., 1980, \auas{42, 251}

\bibitem[1993]{PallaStahler93}
Palla F., Stahler S.W., 1993, ApJ, 418, 414

\bibitem[1998]{Pfeifferetal98}
Pfeiffer M.J., Frank C., Baum\"uller D., Fuhrmann K., Gehren T., 1998, A\&AS,
130, 381
\bibitem[2001]{PrugnielSoubiran01}
Prugniel Ph., Soubiran C., 2001, A\&A, 369, 1048

\bibitem[1992]{Schalleretal92}
Schaller G., Schaerer D., Meynet G., Maeder A., 1992, \auas{96, 269}

\bibitem[1982]{SK82}
Schmidt-Kaler Th., 1982. In {\it Landolt B\"ornstein, New Series, Group IV,
Vol. 2b}, eds. K. Schaifers, H.H. Voigt, Springer, Berlin, Heidelberg, New York

\bibitem[1981]{Swings81}
Swings J.P., 1981, \auas{43,331}

\bibitem[1998]{Zickgraf98}
Zickgraf, F.-J., 1998. In: {\it Dusty B[e] stars}, eds. A.M. Hubert and C. 
Jaschek, Kluwer Academic Publishers, p. 1

\bibitem[1989]{ZS89}
Zickgraf, F.-J., Schulte-\-Ladbeck~R.E., 1989, \aua{214, 274}

\bibitem[1989]{ZickgrafStahl89}
Zickgraf  F.-J., Stahl, O., 1989, \aua{223, 165} (Paper I)


\bibitem[1985]{Zickgrafetal85}
Zickgraf, F.-J., Wolf, B., Stahl, O., Leitherer, C., Klare, G., 1985, A\&A, 143, 
421 

\bibitem[1986]{Zickgrafetal86}
Zickgraf, F.-J., Wolf, B., Stahl, O., Leitherer, C., Appenzeller, I., 1986, 
A\&A, 163, 119 



\end{thebibliography}
\end{document}